# Temporal passing network in basketball: the effect of time pressure on the dynamics of team organization at micro and meso levels

Quentin Bourgeais[a,b]*, Rodolphe Charrier[b], Eric Sanlaville[b], Ludovic Seifert[a]

[a]*University Rouen Normandie, Normandie University, CETAPS UR 3832, F-76000, Rouen, France;*
[b]*Université Le Havre Normandie, University Rouen, Normandie, INSA Rouen Normandie, Normandie University, LITIS UR 4108, F-76600 Le Havre, France;* [c]*Institut Universitaire de France (IUF)*

Corresponding author: quentin.bourgeais@univ-rouen.fr

Quentin Bourgeais http://orcid.org/0009-0004-9623-9565
Eric Sanlaville https://orcid.org/0000-0001-9482-3945
Ludovic Seifert https://orcid.org/0000-0003-1712-5013

Abstract

In this study, basketball teams are conceptualized as complex adaptive systems to examine their (re)organizational processes in response the time remaining to shoot. Using temporal passing networks to model team behavior, the focus is on the dynamics of the temporal patterns of interaction between players. Several metrics grounded in social network analysis are calculated at different level to assess the dynamics of the patterns used by teams and of the individual roles within those patterns. The results reveal a 3-phase dynamic, differentiated by more or less complex and diversified patterns, and by more or less specialized or flexible roles. Additionally, time-dependent features of the different tactical playing positions are identified, some of which linked to team performance. The findings are intended to explain how basketball teams adapt their organization to cope with time pressure, offering potential insights for other type of teams facing similar constraints. Moreover, this work provides a useful framework for a multi-level understanding of how constraints shape team adaptations dynamically, making it applicable to a wide range of team settings.

Keywords
complex adaptive systems; social network analysis; temporal networks; team sports

Highlights
- We provide a multi-level practical framework to analyze how teams adapt their organization under constraints like time pressure
- Time pressure shapes interaction patterns: from simpler/stereotyped, to complex/diverse, then simpler/stereotyped again
- Time pressure shapes player roles: from specialized, to flexible, then specialized again – with variation between positions

**Introduction**

*Conceptual framework*

A *team* is not simply the aggregation of its members. This applies to a wide range of small groups formed for various purposes: a group of workers sharing a project, friends looking to have fun, or a sports team intent on beating its opponent. Such groups cannot be adequately understood as collections of independently acting individuals, as they are driven by interactions both among group members and between the group and its embedding contexts: they should rather be seen as complex, adaptive, and dynamic systems (Arrow et al., 2000). According to Ramos-Villagrasa et al. (2018), the concept of *complex adaptive systems* (CAS) has been applied progressively to organizational sciences and team research, allowing an understanding of such groups, including: groups of workers from small to multinational companies in various fields (*e.g.*, management, academia, healthcare, government, engineering, manufacturing); project groups (*e.g.*, online projects, classroom groups); sports teams (*e.g.*, in football, futsal, rugby, basketball); groups in diverse collaborative tasks or games. The authors argued that the science of teams constitutes a multidisciplinary field, where the CAS framework can contribute toward team scientists speaking a common language (Ramos-Villagrasa et al., 2018).

This conceptual framework is particularly well suited to study team sports, in which a group of individuals share a common goal, collaborate to achieve it, and where the dynamics of their interactions are determinant for team success. As with other groups, a sports team is more than the sum of the players, and research has drawn on insights from organizational sciences to understand why (Eccles & Tenenbaum, 2004). Particularly, Weinberg & McDermott (2002) compared sport and business organizational effectiveness, identifying more similarities than differences in the factors associated with success in both domains (*e.g.*, leadership, cohesion, communication). Theoretical work has also conceptualized sport teams as CAS: Davids (2014) defined them as systems exploiting self-organization processes to achieve task goals by continuously adapting the organization between system components over time, particularly as surrounding constraints change. In the specific case of basketball teams, this approach has proven particularly useful to provide an understanding at different organizational levels, in a dynamical way – particularly through the use of network analyses (Bourbousson et al., 2015).

By essence, using CAS as a framework emphasizes the relationships between group members (Arrow et al., 2000), explaining why previous authors who studied basketball using this framework used network approaches, shedding light on player interactions (*e.g.*, Bourbousson et al., 2015; Fewell et al., 2012). This reflects CAS research applications more broadly, where *social network analysis* (SNA) framework appears to be highly valued as a leading analytical tool (Ramos-Villagrasa et al., 2018). SNA have been used in organizational sciences to study CAS because their focus on relationships enables the understanding of social structures and phenomena that do not exist at the individual level, particularly through the multi-level approach supported by underlying graph theory (Benham-Hutchins & Clancy, 2010). This approach includes: micro-level (*i.e.*, focusing on individual nodes), meso-level (*i.e.*, focusing on subgroups of nodes), and macro-level (*i.e.*, focusing on the entire network). Specifically, the SNA toolkit provides several statistical measures (or *metrics*) to be computed on the network constructed from relational data – see Tabassum et al. (2018) for an overview.

Sport scientists have adapted various of these metrics to study sport team behavior, at the different scales (Clemente et al., 2016), commonly applying them to the passing network (*i.e.*, the network of successful passes between players) of a team over a game (Korte, 2019). This approach has provided insights into team organization, from player roles to team style (Buldú et al., 2019). Notably, Passos et al. (2011) have brought the use of networks in team sports analysis, arguing that the complexity of a team is rooted in the network of interactions between players, and showing that the structure of a team network provides insights about the collective behavior. Also, more than team activity, Duch et al. (2010) showed that analyzing the network of passes is also useful to understand player roles within the team. While such analyses have been predominantly applied to football (Korte, 2019), Fewell et al. (2012) used the same framework to gain insights into player roles and team strategies in basketball.

In present work, SNA framework is used to analyze basketball team organization at different scales, while taking into account a frequently overlooked dimension: time. The interaction network of a system provides an easy way to get an overview of that system, but this is usually done without accounting for the temporal dimension, even though the timing of interactions can have a major impact, thus making it more relevant to model the system as a temporal network (Holme & Saramäki, 2012). In a sports context, Buldú et al. (2019) demonstrated that including the temporal dimension in passing network analysis allows a more in-depth understanding of football team organization. In basketball, time may have an even greater impact due to the *shot clock* rule which limits the time a team can have possession of the ball. According to Skinner (2012), an attacking basketball team can be seen as a system organizing itself to find a good shooting opportunity: players have to choose between shooting or trying to find a better shooting opportunity later in the possession. The time available for the team to shoot (*i.e.*, time remaining on the shot clock) is therefore an important variable in players individual decision-making process, necessarily affecting team organization.

Therefore, from a CAS perspective the time remaining to shoot constitutes a time pressure constraint shaping self-organization processes. In organizational sciences, time pressure has been studied as a team stressor, defined as an environmental event or condition the exposure to which may cause changes in team members capabilities to interact interdependently and achieve the team goal together (Liu & Liu, 2018). Additionally, previous studies have shown that time pressure can affect *team structuring* (*i.e.*, roles distribution within the team). In Isenberg (1981) time pressure is shown likely to cause an increased role-differentiation within small groups, and in Drach-Zahavy & Freund (2007) authors found that having a more mechanistic team structuring (*i.e.*, a team with more specialized roles in comparison to a more organic one with more flexible roles) block the negative effects of increased quantitative stress such as time-pressure. Applied to basketball, it means that the time remaining to shoot might affect the network of passes between the players, particularly the role of each tactical playing positions (*i.e.*, PG: Point Guard, SG: Shooting Guard, SF: Small Forward, PF: Power Forward, C: Center) and its specialization/flexibility. In fact, previous authors have emphasized that basketball team may actually adopt different strategies to organize player roles – *e.g.*, capitalizing on their specialization by moving the ball towards the shooting specialists, or adopting a more flexible approach by distributing the ball to reduce predictability – which was expected to be reflected in the passing network (Fewell et al., 2012).

Yet, the effect of time pressure on basketball team (re)organization processes has not been extensively studied. In particular, an analysis of the effect of the time remaining to shoot on the dynamics of passing interactions and on player roles appear to be worthwhile.

*Network analysis*

Until recently, most passing network studies have focused on static analyses, calculating metrics on aggregated networks at the level of one game (or more). In basketball, Clemente et al. (2015) used classical centrality metrics (*i.e.*, in-degree, out-degree) to analyze the contribution of the different tactical playing positions. They found statistical differences between PG and other positions, showing that PG has a prominent role in the network, meaning that he links team-members and organizes the attacking process (Clemente et al., 2015). This study provides a baseline for centrality levels of the different positions in basketball, however, applying some of the classical network analysis measures to such aggregated data may lead to misinterpretations (Korte, 2019).

In fact, there is a need to consider actual passing trajectories by focusing on the play scale of analysis, that is when a team is in possession of the ball and moves toward the opponent goal (Ramos et al., 2017). For example, in Duch et al. (2010) they constructed the ball flow to evaluate different paths of passes among football teams. Similarly, in Fewell et al. (2012) they analyzed each basketball play as a flow, also accounting for ball income and outcome. In both studies, authors used a *flow centrality* metric to evaluate players centrality in the flow of passes, thus considering the actual passing sequences. Likewise, Korte et al. (2019) introduced a *flow betweenness* metric to evaluate players actual intermediary role (*i.e.*, in the actual flow of passes). Because these metrics are based on actual passing trajectories rather than aggregated data, they are particularly well-suited for evaluating player roles while taking the temporal dimension into account.

Beyond micro scale analyses, the use of ball flow also supports meso scale analyses. But, while micro (*i.e.*, player) and macro (*i.e.*, team) levels are perfectly identified, the meso-level is more subject to further questioning (Buldú et al., 2018). For instance, Wasserman & Faust (1994) include in this intermediate scale the analyses of dyads, triads, or other subsets, a definition implemented in Bourbousson et al. (2015) to analyze the meso-level organization of basketball teams. This level can also be studied by quantifying overrepresented motifs as defined in Milo et al. (2002). This has been investigated in Gyarmati et al. (2014), where authors introduced flow motifs to characterize statistically significant pass sequence patterns. This approach allows the identification of patterns of interaction within subsets of players, thus providing a way to analyze sport teams at meso-level. It has been used in football to identify team playing style (*e.g.*, Gyarmati & Anguera, 2015; Meza, 2017) as well as player roles (*e.g.*, Bekkers & Dabadghao, 2019; López Peña & Sánchez Navarro, 2015), but no such study seems to have been conducted yet in basketball.

**Objectives, hypotheses, and contributions**

*Objectives and hypotheses*

The general aim of this work is to investigate the dynamics of team organization through a multi-level approach combining the micro and meso levels. In particular, the study focuses on the dynamical evolution of: 1) the temporal patterns of passes according to the time remaining on the shot clock, and 2) the role of the different tactical playing positions on those patterns. In order to evaluate the effect of time pressure, these dynamical evolutions are considered in relation to the time remaining to shoot for the offensive.

More precisely, the following three hypotheses were tested. At meso-level: time pressure affects the temporal patterns of passes between the players (Hypothesis No. 1). At micro-level: the Point Guard has a central role in the temporal network of passes (Hypothesis No. 2), and player roles are affected by time pressure, moving from more flexible to more specialized as time pressure increases (Hypothesis No. 3).

*Contributions*

First, this work offers a novel tool for the analysis of social networks, providing a way to integrate the temporal dimension at both micro and meso levels. Actually, it constitutes an extension of the Temporal Passing Network Model (TPNM) developed in a previous work (Bourgeais et al., 2024). In particular, this study delves deeper into the dynamics of relational data by examining the direct effect of time on interaction patterns, introduces the micro-level of analysis to evaluate the role of individual entities within those patterns, and proposes several metrics allowing a finer evaluation of individual and collective behaviors. Although this study is conducted on basketball teams, a core concern of the approach is to ensure that the methodology can be easily applied to any other type of team. Therefore, this work is intended to contribute to the SNA framework, particularly in contexts involving real-time interactions, where time matters in the understanding of the system and its adaptation to its environment.

Second, this work aims to contribute to the interdisciplinary understanding of team (re)organization processes under time pressure conditions, beyond the specific context of basketball teams. Time pressure is a common factor in various domains, which can significantly influence how team members interact and distribute roles – particularly in terms of specialization/flexibility. Thus, the findings are expected to improve the understanding of the way a basketball team works, but also to be applicable across different team settings in which time constraints shape team dynamics. In fact, by focusing on interaction patterns and individual roles, this study addresses questions relevant to any field where teams must adapt quickly under high-stress conditions to optimize their performance.

Together, these two innovations also represent a significant contribution to sport sciences. Since SNA applied to team sports have almost exclusively focused on football, this work is advancing the fundamental understanding of basketball teams. It follows in the footsteps of previous authors who have studied basketball teams through network analyses, in particular Fewell et al. (2012) in which they used the flow of the ball within the team to identify team strategies and player roles, and Bourbousson et al. (2015) in which they proposed a multi-level analysis of the patterns of social interactions between players. Moreover, this work contributes to the field of performance analysis in team sports by integrating the temporal dimension into passing network analysis, something recently recommended and/or investigated by several authors (*e.g.*, Buldú et al., 2018; Korte et al., 2019; Ramos et al., 2017). In the end, this work could provide valuable insights into performance optimization in basketball: by providing insights into temporal patterns of interaction and player roles within basketball teams, it may support the development of coaching strategies aimed at enabling more adaptive team organizations, particularly when facing high-pressure conditions.

**Materials and Method**

This section presents: 1) the TPNM designed to characterize the dynamics of passes between players; 2) the data used to conduct the study; 3) the network and statistical analyses performed.

*Temporal Passing Network Model*

The TPNM extensively described in Bourgeais et al. (2024) has been applied to capture the dynamics of passing interaction between basketball players during a possession. This model is based on graph theory: a graph is defined as a pair $G = (V, E)$ where $V$ is a set of vertices and $E$ a set of edges between these vertices. The main feature of the TPNM is the use of a sliding time window from the beginning to the end of each possession. The time window has two parameters: a duration (*i.e.*, difference in time between the end and the start of the time window) and a step (*i.e.*, difference in time between the start of two consecutive time windows). Here, a 6-s duration and a 0.5-s step are used, meaning that a time window ($T_k$) runs from $t_k$ to $t_k + 6s$ and that the next time window ($T_{k+1}$) starts at $t_k + 0.5s$. Therefore, time is discretized into a succession of time windows, and each time window $T_k$ is associated with a graph $G_k = (V_k, E_k)$ where $V_k$ and $E_k$ are respectively the sets of vertices and edges that exist during $T_k$ (*i.e.*, the set of successful passes and the set of players involved within the time interval $t_k$ to $t_k + 6s$). The graph $G_k$ is called a *snapshot*, and each possession is modeled as a *temporal graph* named $G$, defined as the sequence of all snapshots $\{G_1, G_2, …G_n\}$.

*Data*

The dataset includes all 12 games of the 2019 Men's FIBA Basketball World Cup from the quarter finals. Conducting the study over this sample ensures certain redundancy and homogeneity in the data: it involves 8 teams, each playing 3 games, against 3 of the 7 other teams, and this in a very similar context. Video footage of each of these games, available on the FIBA's YouTube channel, has been manually annotated by one observer (using Dartfish software) to obtain information on each successful pass, including timecode, passer, and receiver. Additionally, the possession starting and ending timecodes, the time on the shot clock and the ball location at the start of the possession, and the final outcome of the possession have been annotated. An intra-rater relatability analysis has been conducted to evaluate the validity of measured timecodes by repeating the manual annotation of pass events on a randomly selected subset representing 10% of the total dataset: it revealed that event timecodes differ by less than 0.25s in 96.4% of cases.

The raw data contains 2,213 possessions for a total of 6,073 successful passes. Possessions shorter than 6s have been excluded because their duration is shorter than a single time window: once this filter is applied, 1,751 possessions remain (*i.e.*, 20.9% of the possessions were excluded) for a total of 5,554 passes (*i.e.*, 8.5% of the passes were excluded). The use of this 6s filter is also intended to ensure that fastbreak possessions are removed from the data, so that the study has a focus on set plays. In order to compare situations from the same tactical category even more precisely – because it might affect the impact of the remaining shot clock – those possessions have been then divided into categories based on their starting location on the court. As a result: 998 possessions started with a ball offside in the defensive half, 537 started with a ball onside in the defensive half, 157 started with a ball offside in the offensive half, and 59 started with a ball onside in the offensive half. In view of the amount of data available, only the first two categories (*i.e.*, those starting in the defensive half) have been retained for analysis, representing 87.7% of the possessions that lasted 6s or more (*i.e.*, 1,535 possessions). Whitin the remaining possessions, two possession types are therefore defined: "ball in" (*i.e.*, those starting with an inside ball) and "ball out" (*i.e.*, those starting with an offside ball). Finally, running the time window procedure over the 1,535 possessions resulted in 24,908 single time windows – and so 24,908 snapshots.

*Network and statistical analyses*

*At meso-level*

A first analysis investigates the effect of the shot clock remaining on the temporal passing patterns between players. To this aim, only the network structure is retained from each snapshot, thus disregarding the individuals involved – as for flow motifs. In the model, the concept of *graphlets* (Pržulj et al., 2004) is used rather than motifs, to be more in line with the literature on subgraphs – see Bourgeais et al. (2024).

The approach consists in listing possible graphlets and then classifying each snapshot as one of them. Only the patterns with up to 3 passes are considered, giving a total of 9 possible graphlets (Figure 1). A 10th category is added for cases in which the structure of the snapshot does not match any of the 9 graphlets (*i.e.*, when there are 4 passes or more). Then, the purpose is to create a *graphlet profile*, defined as the collection of frequencies of each graphlet, to characterize temporal passing networks by aggregating possessions according to a given criterion (*e.g.,* all possessions of a given team).

**Figure 1**

*List of all the graphlets*

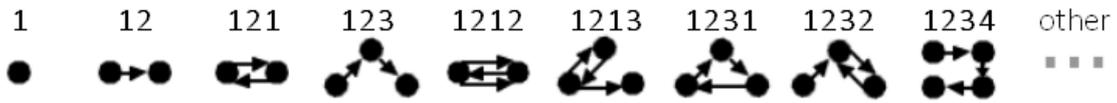

Note. This figure represents all graphlets considered: "1" to "1234" are the 9 possible graphlets with 0 to 3 edges, and "other" contains all graphlets with at least 4 edges.

Each snapshot is associated with its starting shot clock value (*i.e.*, the time remaining on the shot clock at the beginning of the time window), ranging from 24s (*i.e.*, the maximal value because of the rules of the game) to 6s (*i.e.*, the minimal value because this is the duration of a time window). A graphlet profile is created for each possible starting shot clock value, resulting in 37 profiles – called *shot clock graphlet profiles*.

In addition, the *State Entropy* (SE) metric used in Bourgeais et al. (2024) is calculated on all shot clock graphlet profiles. SE corresponds to the classical definition of entropy for a distribution (Shannon, 1948). In the context of this work, it informs on the use of various temporal patterns of interaction: the higher the entropy, the greater the diversity of the patterns used. This procedure is repeated twice: once for "ball in" possessions, and once for "ball out" possessions.

Several $\chi^2$ independence tests are performed to compare shot clock graphlet profiles, following this procedure: considering the shot clock value $s$ ranging from 24s to 6s with a -0.5s increment, the graphlet profile $GP_s$ associated to the shot clock value $s$ is first compared to the graphlet profile $GP_{s-0.5}$ associated to the following shot clock value $s - 0.5$; if there is no significant difference between their distributions, the graphlet profile $GP_s$ is then compared to the next graphlet profile $GP_{s-1}$ associated to the next shot clock value $s - 1$; and this operation is repeated as many times as necessary to identify a significant difference – if any. Hypothesis No. 1 is tested using the results of these tests, searching for statistical differences between the shot clock graphlet profiles.

*At micro-level*

A second analysis focuses on the role of individuals in the network, grouped by tactical playing position (*i.e.*, PG, SG, SF, PF, C) based on information available on the official competition website. As each individual player is assigned to a tactical playing position a priori, it means that the same position can be on the court several times at once – or conversely, a position can be away from the court. This micro-level analysis is composed of two parts.

In the first part, the purpose is to understand the role of the different positions in the temporal passing network. To do so, the analysis of football player roles conducted in Korte et al. (2019) is reproduced, using the same flow-based metrics: flow centrality (FC) and flow betweenness (FB). FC is defined as the fraction of possessions in which a given player is involved, and FB the fraction of possessions in which a given player is actually in-between other teammates in the passing sequence (*i.e.*, when he receives the ball and then gives it to a teammate). Thus, it starts by calculating the two classical flow-based metrics for each position and reproducing the statistical analysis conducted in Korte et al. (2019) to identify any differences in positions (*i.e.*, Kruskal–Wallis *H* tests and non-parametric estimates of $\eta^2$ with their 95% confidence interval). Hypothesis No. 2 is tested in this way.

Following the methodology of previous authors, possessions outcomes are classified as either positive or negative according to their outcome: positive outcomes include scoring points and/or provoking opponent fouls, while conversely a possession has a negative outcome if it ends with no points scored and/or no opponent fouls. But then, to compare successful and unsuccessful possessions, a different test from the original study is performed. As flow-based metrics consist in attributing either a 0 or a 1 for each possession, it can be considered as a proportion of 0 and 1 among a given number of possessions. Thus, $\chi^2$ independence tests are suited to search for statistical differences in these metrics between the different possession outcomes. These tests are computed for each position, and for both types of possessions. In addition, the measure of difference between the two metrics for each playing position is not computed: because of the huge differences between playing positions in terms of FC, the use of a ratio (*i.e.*, FB over FC) seems more appropriate.

The same analysis is conducted a second time, but using flow-based metrics adapted to a graphlet-level: this time, it is the fraction of graphlets in which a given individual is involved or in-between other individuals in the passing sequence that is calculated. These *adapted metrics* are thought to measure in a finer way the contribution of each individual to their team's network, by taking the temporal dimension more fully into account. Indeed, the actual involvement or intermediary role of an individual in a possession can be very different whether it occurs only in one graphlet or in all graphlets of that possession. For instance, if that possession has 10 graphlets, that individual's score would be 0.1 in the first case (*i.e.*, 1 out of 10) and 1 in the second (*i.e.*, 10 out of 10) at graphlet-level, whereas at play-level it would be the same score of 1 in both cases. This illustrates how a same score on an original flow-based metric can actually conceal a wide range of scores if calculated at graphlet-level.

Then, the same $\chi^2$ independence tests are computed again with the adapted metrics to search for statistical differences between the different possession outcomes, for each position and for both types of possessions.

Additionally, the relationship between the original metrics and the adapted metrics is examined using regression analysis. To this aim, adapted metrics need to be aggregated over a game (*i.e.*, mean and 95% confidence interval) so that they can be compared with the original metrics, at game-level. This is also repeated twice: once for "ball in" possessions, and once for "ball out" ones.

In the second part, the purpose is to evaluate the how player roles evolve according to the time remaining to shot. First, adapted flow-based metrics are calculated, for each shot clock value, as the fraction of graphlets in which a given individual is involved or in-between other individuals among all the graphlets sharing the same shot clock value. This is repeated for each position, each time resulting in a score of adapted FC and adapted FB (*i.e.*, mean and 95% confidence interval).

Then, the graphlet profile of each position is generated for each shot clock value. Each of these profiles – called *individual shot clock graphlet profiles* – therefore corresponds to the collection of frequencies of each graphlet associated to a given shot clock value and in which a given position is involved. This is done separately for "ball in" and "ball out" possessions.

For each individual shot clock graphlet profile, SE is calculated to evaluate whether the player's role is more flexible or specialized: the higher the entropy, the greater the diversity of patterns in which the individual is involved or in-between other individuals, the more flexibility the player's role has at this specific shot clock value – and vice versa. Hypothesis No. 3 is tested in this way.

**Results**

*Description of the data*

Graphlet profiles (Figure 2A) indicate that the most common graphlet is "12", followed by "1", "123", and "121". Other graphlets are less frequent, but "1234" is prominent among the rarer graphlets. The graphlet "1212" is the least used. This is true for both possession types, and there are only minor differences in their distributions: 2.13% for graphlet "1", 1.82% for graphlet "12", and less than 1% for all others. The number of time windows in the data shows a gradual decrease as the shot clock value decreases (Figure 2B). This results from the procedure used: existence of the time window $T_{k+1}$ implies the existence of the time window $T_k$, but not vice versa. The fastest decrease of the number of time windows for "ball in" possessions indicates that there are more possessions ending with more time remaining. In the dataset, 764 possessions (49.8%) have a negative outcome and 708 possessions (46.1%) have a positive outcome (Figure 2C). In particular, for "ball in" possessions the proportion is 49.0% negative and 47.5% positive, while for "ball out" it is 50.2% negative and 45.3% positive. Comparing these proportions using a $\chi^2$ test reveals no significant difference [$\chi^2$ (df = 1, N = 1472) = 0.34, p = 0.559] meaning that the outcome repartition is similar for both types of possessions.

**Figure 2**

*Descriptive representation of the dataset*

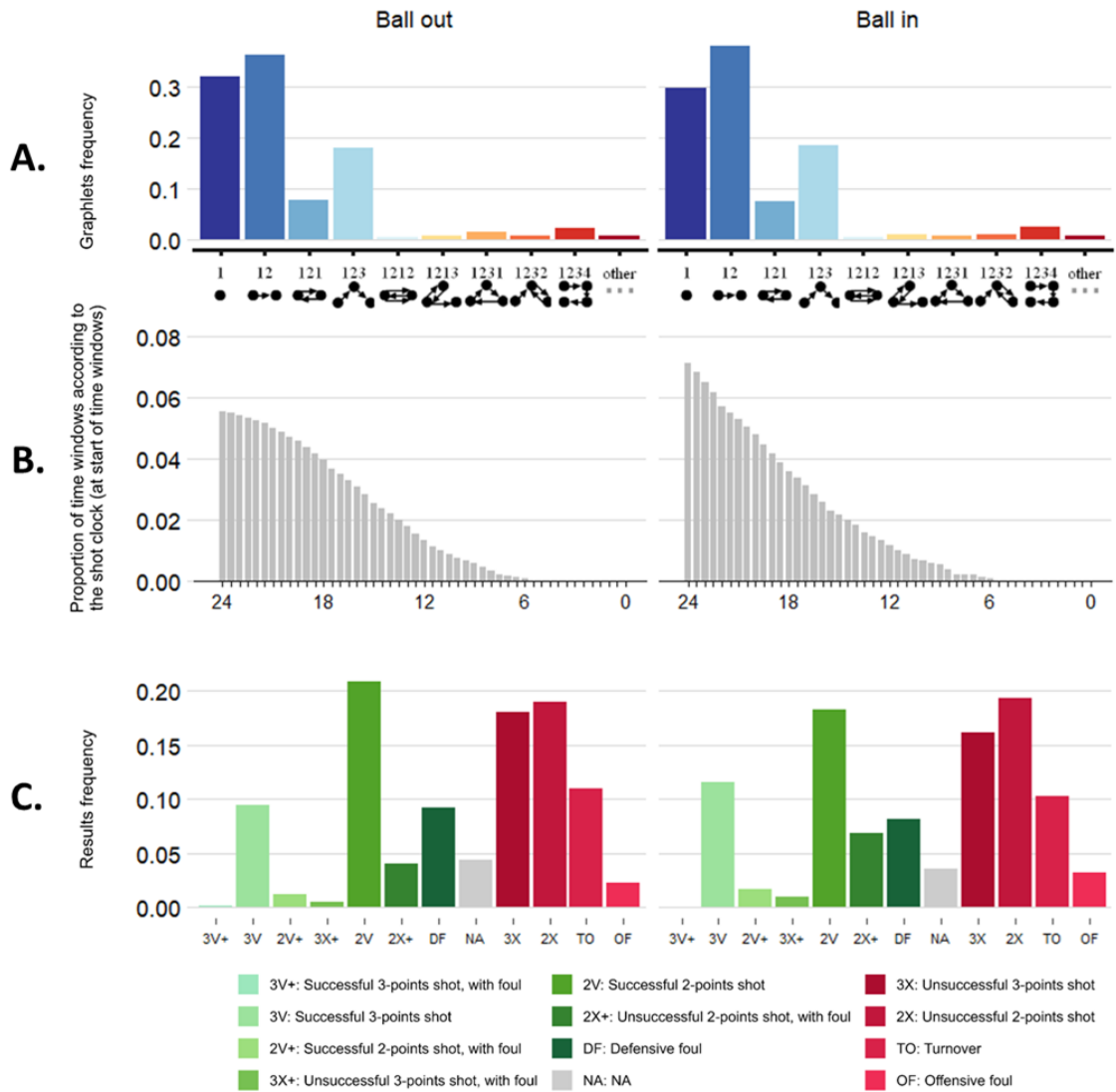

*Note.* This figure represents the dataset, split by possession type: "ball out" (left panels) and on "ball in" (right panels). Each time window is associated with a graphlet and a shot clock value (*i.e.*, the shot clock remaining at the beginning of the time window), and each possession is associated with an outcome.

[A] Frequency distribution of graphlets, as a frequency. [B] Distribution time windows associated to each possible shot clock value. [C] Distribution of the possession outcomes (positive outcomes in green, negative outcomes in red).

*Meso-level analysis*

Shot clock graphlet profiles show an evolution of the graphlets according to the time remaining to shoot (Figure 3A). Several observations can be made based on the graphlet frequencies over time, the main ones are the following. The first time window of "ball-out" possessions reveals a very unique graphlet distribution, characterized by the absence of graphlet "1" (*i.e.*, a player carrying the ball during the entire time window) in comparison to other shot clock graphlet profiles. This is because a "ball-out" possession requires an incoming pass to start. But then, for these possessions, this graphlet "1" occurs in more than 50% of the first time windows and rapidly decreases to stabilize around 20% before increasing slightly when shot clock values are low. The same dynamics are observed for this graphlet "1" within "in-ball" possessions, but with some differences: it never reaches the same frequency (*i.e.*, never more than 40%), it requires several time windows before to reach its maximal frequency, and it remains stable at a higher level (*i.e.*, around 25%). In contrast, the pattern "12" (*i.e.*, a player passing the ball to a teammate) occurs in more than 50% of the first time windows in "ball-in" possessions, and rapidly decreases with the increase of graphlet "1". The dynamics of the graphlet "1" also allow all other graphlets to start with a higher level in "ball-in" possessions than in "ball-out" ones: this is particularly true for graphlet "123", which is the third more prominent pattern of interaction in most of 6s time windows. Mirroring the dynamics of the three main patterns of interaction (*i.e.*, "1", "12" and "123"), the more complex ones (*i.e.*, those with at least 3 edges) have lower frequencies when the shot clock value is high or low, but greater frequencies in the intermediate ones (*i.e.*, reaching around 10% when summed up). These complex patterns are therefore more likely to appear after several time windows, and are likely to disappear when shot clock values are low. It also can be observed that graphlets with 4 edges or more tend to occur more in "ball-in" possessions than in "ball-out" ones, and that their occurrence comes even later than graphlets of 3 edges.

**Figure 3**

*Shot clock graphlet profiles and associated level of entropy*

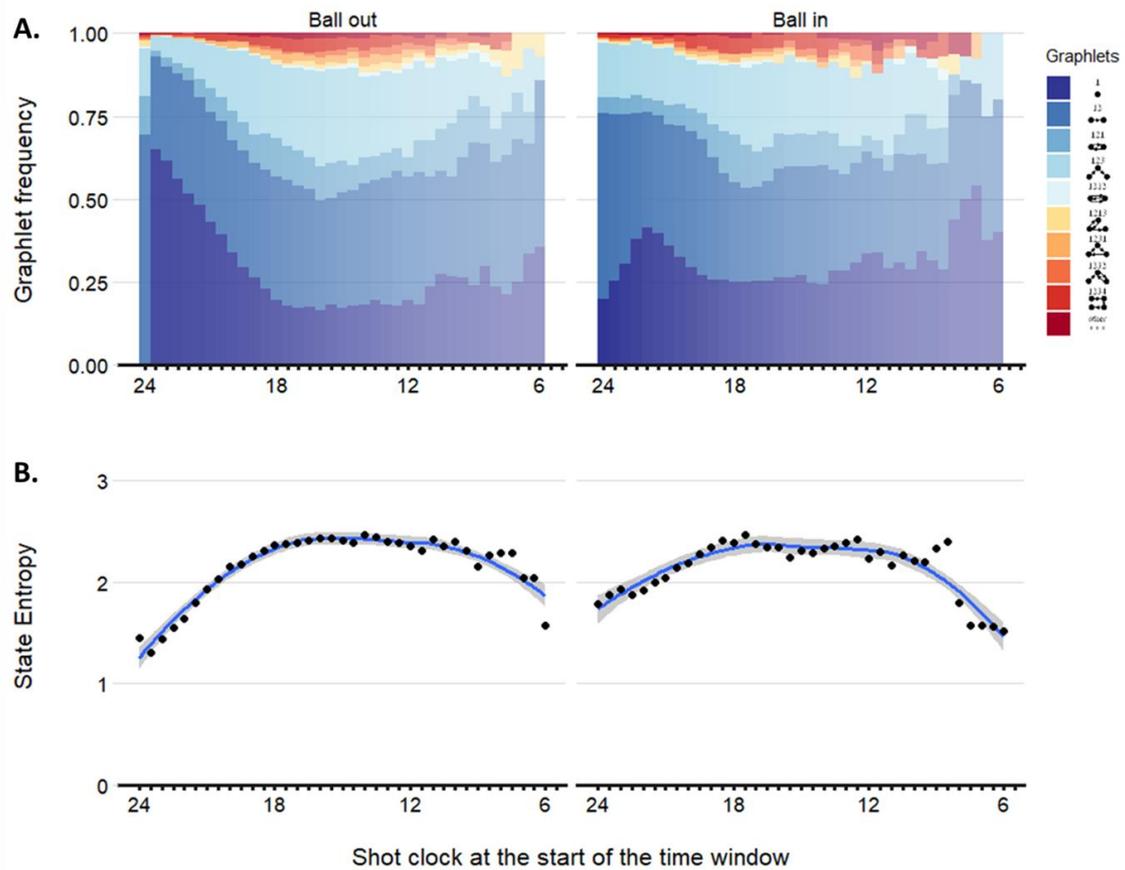

*Note.* This figure represents the shot clock graphlets profiles and their level of entropy, according to their associated shot clock value, split by possession type: "ball out" (left panels) and "ball in" (right panels).

[A] The collection of graphlet frequencies for each shot clock value at the start of the time window (*i.e.*, shot clock graphlet profiles), calculated as the number of occurrences of a given graphlet among all the graphlet sharing the same shot clock value. [B] The State Entropy (*i.e.*, SE) calculated on all shot clock graphlet profiles: black points represent SE, blue line is the smooth function (using Generalized Additive Model) with its 95% confidence interval in grey.

This evolution of graphlets over time also leads to changes in the level of entropy in shot clock graphlet profiles (Figure 3B). At the highest shot clock values, the level of entropy is low and increases progressively to reach its maximal values when the remaining shot clock is around 18s. This peak of entropy is comparable for "ball-in" and "ball-out" possessions, but "ball-out" possessions start at a lower level than "ball-in" ones. Once this peak is reached, it remains stable for a large part of intermediate shot clock values, before decreasing again when shot clock values are at their lowest (to slightly lower levels for "ball-in" possessions).

The multiple $\chi^2$ independence tests computed to compare the shot clock graphlet profiles statistically mainly supports this 3-phase dynamic as a function of the shot clock value: an initial evolving phase for high values, a stable phase for intermediate values, and a different ending phase for the lowest values (Figure 4). Indeed, results reveal that for a shot clock value between 24s and 19.5s or 18.5s (*i.e.*, 19.5s for "ball-in" and 18.5s for "ball-out" possessions) any shot clock graphlet profiles have a different one within the following ones. This means that the graphlet distribution is changing during that phase, with high shot clock values. Then, for graphlet profiles corresponding to a shot clock value of 19s and 18s respectively, the next one different is at shot clock values of 10.5s or lower, thus with no significant difference in-between. This means that the graphlet distribution remains stable during that phase associated to intermediate shot clock values. Finally, the same thing appears for shot clock values of 10.5s and 11s respectively for "ball-in" and "ball-out" possessions, but this time there is no next one different since the number of remaining shot clock values is limited. This means that there is a new graphlet distribution for these low shot clock values.

**Figure 4**

*Shot clock graphlet profiles comparisons*

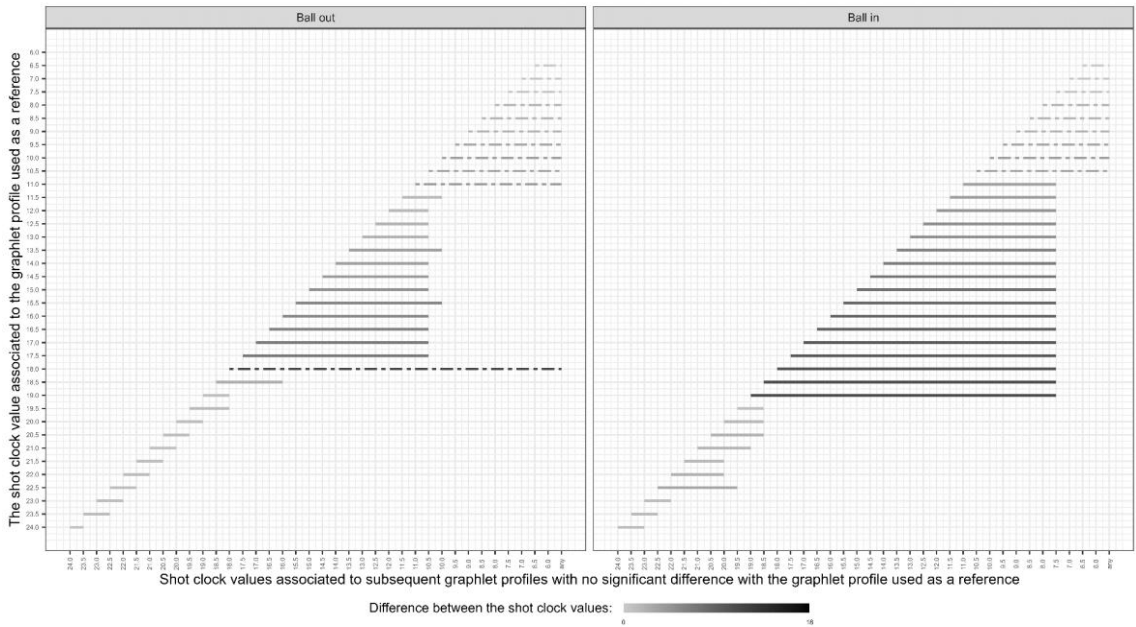

*Note.* This figure represents the results of the comparisons of shot clock graphlet profiles following the procedure explained in Materials and Method section, for both types of possession: "ball out" (left panels) and "ball in" (right panels). Horizontal bars represent the distance between the shot clock values associated to a shot clock graphlet profile and the first significantly different shot clock graphlet profile – if any (i.e., the longer the bar, the greater the number of similar consecutive profiles).

*Micro-level analysis*

The reproduction of the analysis in Korte et al. (2019) shows significant differences between positions, with moderate effect on FC (for "ball-in": $p < 0.001$; $\eta^2 = 0.37$, CI [0.34, 0.41]; for "ball-out": $p < 0.001$; $\eta^2 = 0.43$, CI [0.40, 0.45]) and a small effect on FB (for "ball-in": $p < 0.001$; $\eta^2 = 0.17$, CI [0.14, 0.20]; for "ball-out": $p < 0.001$; $\eta^2 = 0.19$, CI [0.17, 0.21]). The most central player is the PG, followed by the SG, SF, C, and PF in that order (Figure 5), for both FC and FB, and for both types of possessions.

The comparison of the level of FC and FB between successful and unsuccessful possessions shows no effect for any playing position, meaning that there is no link between the role of different playing positions and the possession outcome for neither type of possessions.

**Figure 5**

*Comparisons of original flow-based metrics according to the possession outcome*

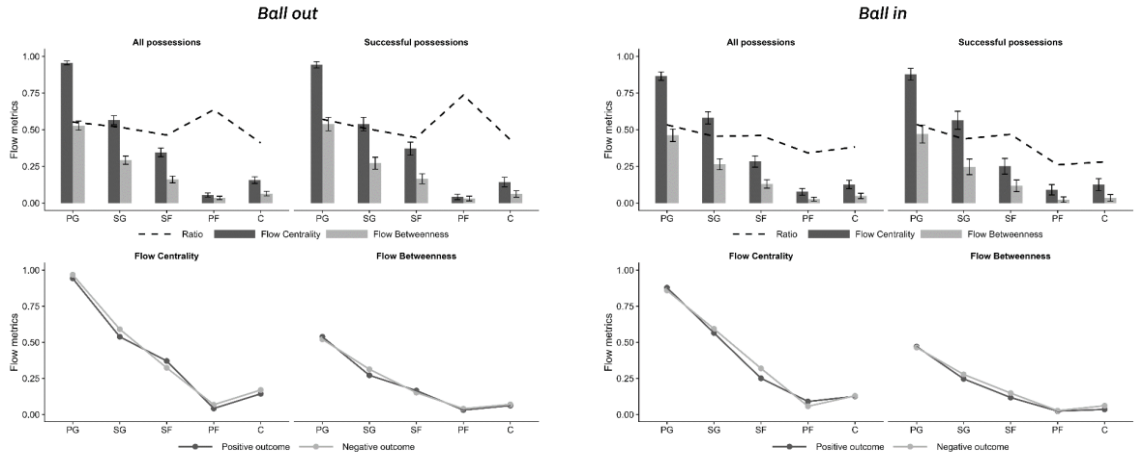

*Note.* This figure represents original flow-based metrics (*i.e.*, mean FC and FB calculated at play-level, with 95% confidence intervals for upper figures) by playing position, according to the possession outcomes, repeated for both types of possessions: "ball out" (left panels) and "ball in" (right panels). This is a reproduction of Figure 2 in Korte et al. (2019), but with ratio between both metrics (*i.e.*, FB over FC) instead of difference.

The same tests repeated using FC and FB adapted to graphlet-level also show significant differences between positions, with comparable effects: moderate effect on adapted FC (for "ball-in": $p < 0.001$; $\eta^2 = 0.42$, CI [0.41, 0.43]; for "ball-out": $p < 0.001$; $\eta^2 = 0.46$, CI [0.46, 0.47]) and small effect on adapted FB (for "ball-in": $p < 0.001$; $\eta^2 = 0.06$, CI [0.06, 0.07]; for "ball-out": $p < 0.001$; $\eta^2 = 0.06$, CI [0.06, 0.07]). The player hierarchy remains the same in terms of centrality and betweenness, for both types of possessions.

The comparison of the level of adapted FC and adapted FB between successful and unsuccessful possessions shows some significant effects, that differ by playing position and possession type, meaning that there are some links between the role of some playing positions and the possession outcome. For "ball-in" possessions, there are significant differences for adapted FC in SG position [$\chi^2$ (df = 1, N = 7100) = 14.34, p < 0.001, OR = 1.20, CI [1.09, 1.32]] and SF position [$\chi^2$ (df = 1, N = 7100) = 6.21, p = 0.012, OR = 1.18, CI [1.04, 1.34]], and for adapted FB in SG position [$\chi^2$ (df = 1, N = 7100) = 5.32, p = 0.020, OR = 1.20, CI [1.03, 1.41]] and PF position [$\chi^2$ (df = 1, N = 7100) = 4.13, p = 0.039, OR = 1.62, CI [1.02, 2.61]]. In comparison to unsuccessful possessions, results indicate that in successful possessions: SG position is less involved and less intermediate (*i.e.*, a 11.0% smaller adapted FC; a 18.2% smaller adapted FB), SF position is less involved (*i.e.*, a 14.4% smaller adapted FC), and PF is less intermediate (*i.e.*, a 60.8% smaller adapted FC). For "ball-out" possessions, there are significant differences for adapted FC in PG position [$\chi^2$ (df = 1, N = 16868) = 4.74, p = 0.029, OR = 1.11, CI [1.01, 1.22]] and PF position [$\chi^2$ (df = 1, N = 16868) = 27.49, p < 0.001, OR = 1.67, CI [1.38, 2.04]], and for adapted FB in PG position [$\chi^2$ (df = 1, N = 16868) = 12.74, p < 0.001, OR = 0.87, CI [0.80, 0.94]] and PF position [$\chi^2$ (df = 1, N = 16868) = 6.68, p = 0.009, OR = 1.50, CI [1.10, 2.06]]. In comparison to unsuccessful possessions, results indicate that in successful possessions: PG position is less involved but more intermediate (*i.e.*, a 1.3% greater adapted FC; a 11.5% greater adapted FB), and PF position is less involved and less intermediate (*i.e.*, a 39.4% greater adapted FC; a 32.9% smaller adapted FB).

Comparing the original flow-based metrics (*i.e.*, calculated at the play-level) with the adapted flow-based metrics (*i.e.*, calculated at graphlet-level) shows nonlinear and non-proportional relationships (Figure 6). For FC, the difference is maximal for intermediate values and decreases as values approach 0 or 1, thus impacting particularly the evaluation of the roles of SG and SF. For FB, the difference increases as the FB value becomes higher, thus mainly affecting the evaluation of PG, then SG and SF, and finally PF and C roles to a smaller extent. This is true for both types of possessions.

**Figure 6**

*Comparisons between original flow-based metrics and adapted flow-based metrics*

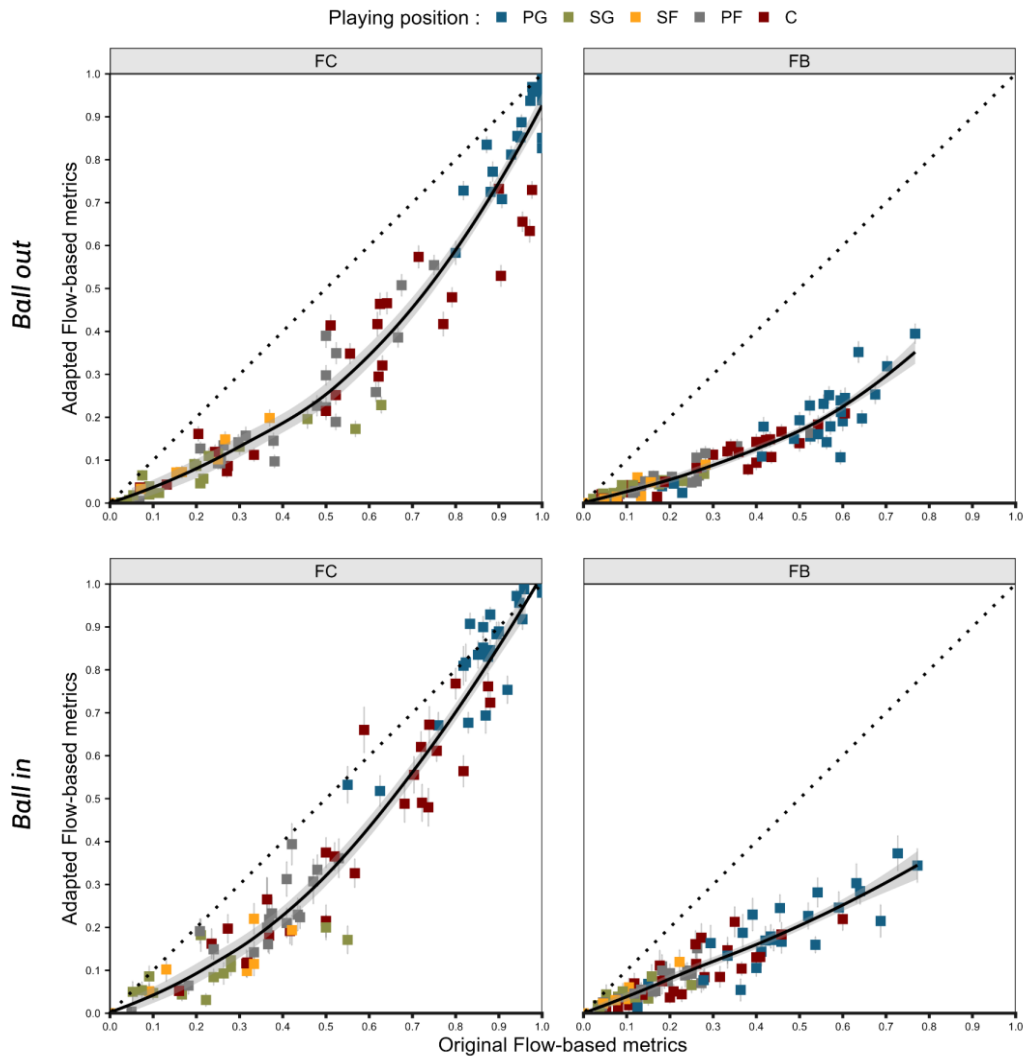

*Note.* This figure represents the comparisons between original mean flow-based metrics (*i.e.*, FC and FB calculated at play-level) and adapted mean flow-based metrics (*i.e.*, FC and FB calculated at graphlet-level). Adapted metrics are aggregated by calculating their average and 95% confidence interval over each game to be compared with the original metrics. Squares represents the value for 1 playing position in 1 game, for FC (left panels) or FB (right panels). Dashed lines show theoretical linear proportional relationships. Solid black lines are the smooth function (using LOESS) with its 95% confidence interval in grey. This is repeated for both types of possessions: "ball out" on the top and "ball in" on the bottom.

The analysis of flow-based metrics adapted to graphlet-level according to the shot clock value shows common and unique features among playing positions dynamics (Figure 7). While the player hierarchy mainly remains the same in terms of centrality and betweenness regardless of shot clock value and type of possessions, the dynamics of adapted FC and FB depend on both the playing position and the type of possessions. The PG position has a unique role compared to other positions, characterized by a higher implication for greater shot clock values than for lower ones. The role of this prominent position changes according to the type of possessions: a relatively high centrality but low betweenness during the initial phase of "ball-out" possessions, compared to a relatively lower centrality but higher betweenness in that phase for "ball-in" ones. The PG position also appears to lose even more importance (*i.e.*, both centrality and betweenness) for the very lower shot clock values in the case of "ball-in" possessions. Other positions share more common dynamics, with an effect of the type of possessions on their centrality, but not on their betweenness. In "ball-out" possessions, SG, SF, PF and C positions all show a relatively low initial centrality that increases progressively as the shot clock value decreases, with C position demonstrating a delayed increase (*i.e.*, reaching its peak at lower shot clock values). In comparison, in "ball-in" possessions, the centrality of all these positions is at its average value even for the higher shot clock values. Some differences in the dynamics of centrality still appear between these positions for the very lowest shot clock values: SG position shows an increase in centrality for "ball-in" possessions and a decrease for "ball-out" ones, and conversely SF and PF positions show a slightly higher centrality in "ball-out" possessions than in "ball-in" ones for these lowest shot clock values. In terms of betweenness, SG, SF, PF and C positions are characterized by a relatively low initial level, increasing for intermediate shot clock values, and decreasing as the shot clock decreases – with C position showing slightly delayed dynamics again.

**Figure 7**

*Evolution of adapted flow-based metrics according to the time window, by position*

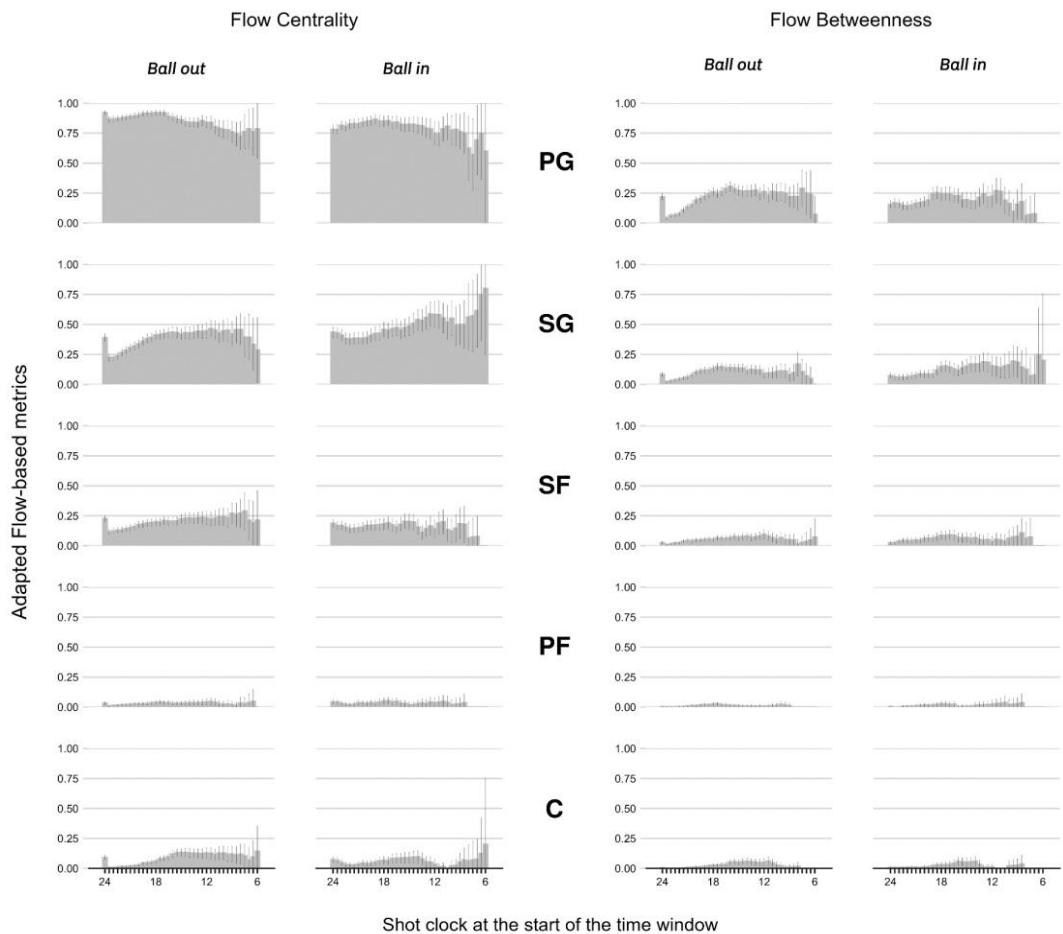

*Note.* This figure represents the evolution of adapted flow-based metrics, by playing position, according to the shot clock value: mean FC (left panels) and mean FB (right panels) calculated at graphlet-level, with their 95% confidence intervals. This is repeated for both types of possessions (for each metric: "ball out" on the left and "ball in" on the right).

The detailed examination of individual shot clock graphlet profiles reveals position-specific dynamics (Figure 8A). For PG position, these profiles closely mirror the macro-level shot clock graphlet profiles, which is consistent with the high score of flow centrality of that position. SG position shows profiles relatively comparable to PG, although with a lower prevalence of graphlet "1", thus allowing for greater proportions of other graphlets. However, there is a distinction between the dynamics of these roles for lower shot clock values: the graphlet "1" increases in relative frequency for SG, and this is even more true for "ball-in" possessions. SF, PF, and C positions display very different profiles than PG and SG, mainly characterized by the near absence of graphlet "1". This graphlet only appears at specific shot clock values, but only in "ball-out" possessions: at very highest and very lowest values for SF position, and to a lesser extent at some low values for PF and C positions. Consequently, SF, PF and C positions have proportionally more complex interaction patterns. That is particularly true in "ball-in" possessions, with a differentiate effect of positions: a high proportion of complex graphlets appears at low shot clock values for SF and PF positions, while it appears at high shot clock values for C position.

In terms of entropy, all individual shot clock graphlet profiles shows the same overall dynamics than the macro-level shot clock graphlet profiles, but with position-specific variations in terms of entropy level, timing of transitions, and/or differences between types of possessions (Figure 8B). For high shot clock values, PG, SG, and SF positions show higher entropy levels than PF and C positions, with an exception for the PG position which displays a relatively low diversity of graphlets in "ball-out" possessions compared to "ball-in" ones. Then, for all positions in both types of possessions, the level of entropy increases as the shot clock value decreases, until reaching their highest relative levels around 18s, except for the C position for which it is only around 14s. This high entropy level is maintained longer in "ball-out" possessions for SF, PF and C positions than in "ball-in" ones. In any case, the entropy level drops for all positions and for both types of possessions once shot clock values are low. It means that, after being involved in diverse patterns of interaction, each playing position returns to more specific patterns when the time remaining to shoot is short.

**Figure 8**

*Individual shot clock graphlet profiles and associated level of entropy*

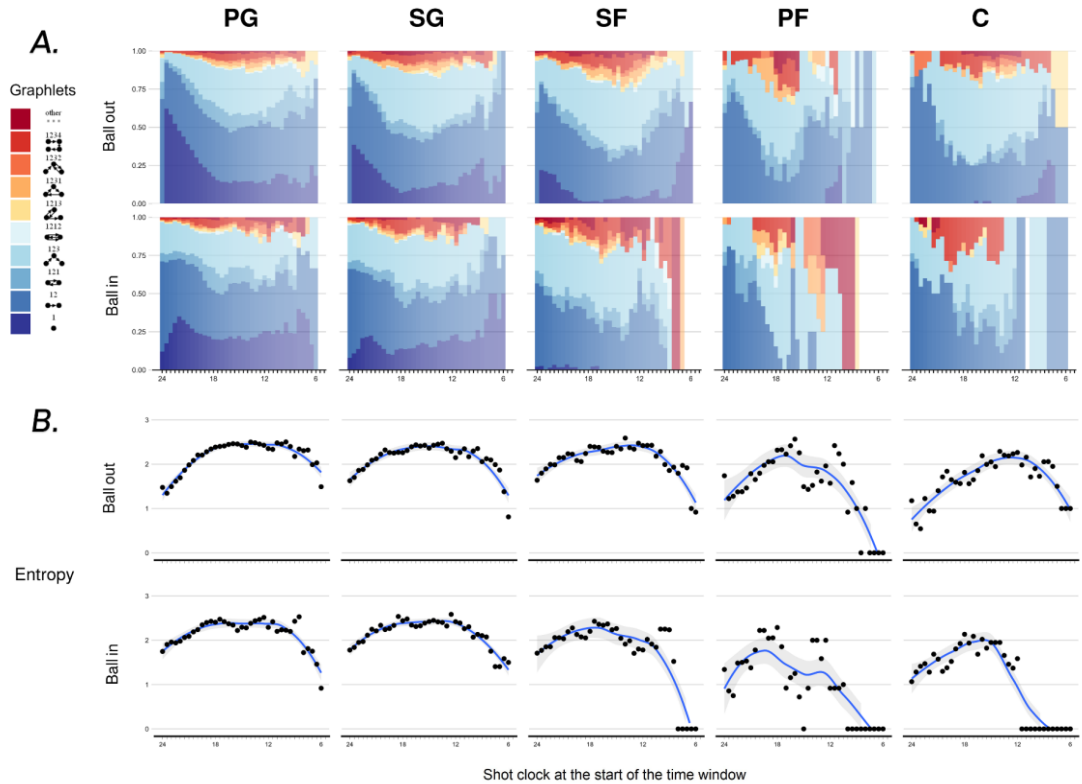

*Note.* This figure represents the individual shot clock graphlets profiles and their level of entropy, according to their associated shot clock value, split by possession type (for both: "ball out" on the top and "ball in" on the bottom).

[A] The collection of graphlet frequencies for each shot clock value at the start of the time window for each tactical playing position (*i.e.*, individual shot clock graphlet profiles), calculated as the number of occurrences of a given graphlet in which a given tactical position is involved among all the graphlet sharing the same shot clock value in which that position is involved. [B] The State Entropy (*i.e.*, SE) calculated on all individual shot clock graphlet profiles: black points represent SE, blue line is the smooth function (using Generalized Additive Model) with 95% confidence intervals in grey.

**Discussion**

*Team organization: dynamics of temporal patterns of passes*

In line with the first hypothesis, changes in graphlet frequencies can be observed according to time window boundaries. This evolution in the temporal patterns of passes between players during set play possessions indicates that team organization evolves according to the time remaining on the shot clock. The detailed analysis of this adaptive behavior reveals a 3-phases dynamic. First, an initial phase, from the ball recovery to around 19s remaining on the shot clock, in which teams are setting up their organization. Obviously, the ball recovery conditions have a major influence on the graphlets appearing during that phase. Second, a stable phase, from approximatively 19s to 11s remaining on the shot clock, in which teams maintain a given organization. It is in that phase that teams exhibit the more diverse and complex behavior. Third, a critical phase, starting when the time remaining to shoot drops below around 11s, in which teams re-organize their behavior. Thus, it is in that last phase that the effect of the shot clock on team organization seems to take shape. From the perspective of a team understood as a complex adaptive system, this reflects the system's adaptation to changes in the time constraint, through changes in interaction patterns among entities.

Consistent with the conception presented in Skinner (2012), the last two phases constitute the core of the attacking play, in which the offensive team tries to create shooting opportunities – and eventually select one of them. This can last quite a long time, until a critical point is reached, at which it is no longer likely to find a better shooting opportunity later in the possession: therefore, when time pressure becomes critically high, players stop searching for another one and opt to exploit their own – however slight. Results indicate that this critical value is around 11s remaining on the shot clock. Interestingly, team organization dynamics are shared among possessions which do not belong to the same tactical category. Although there are differences in the graphlets used, mostly during the initial phase, the shot clock values separating the three phases are very similar. This is probably explained because the situations involved were not so different, apart from the very firsts time windows – and more particularly the first one.

*Player roles: centrality and betweenness*

The micro-level analysis confirms the second hypothesis, with results consistent with previous research. The PG position has a major contribution on team organizational processes, both in terms of centrality and betweenness, and this consistently no matter the shot clock value. It reveals a role that is both highly involved and important to link the other playing positions. SG and SF also play significant roles, although to a lesser extent, while C and PF are less involved in the temporal passing network. However, the difference between positions is more pronounced than in previous studies, especially when compared to Fewell et al. (2012). This may partly be due to differences in the datasets (*i.e.*, 2010 NBA playoffs vs. 2019 FIBA World Cup) but mainly by the ways used to assign positions: they used actual starting five players (and substitutions) to assign each tactical playing positions to a player on the court, while present work is based on a players' own characteristics (*i.e.*, using a priori player official playing position). This methodological choice ensures to fit more with actual basketball team strategies, not assuming a constant theoretical configuration (*i.e.*, 1 PG + 1 SG + 1 SF + 1 PF + 1 C) but allowing every possible team composition (*e.g.*, 2 PGs + 1 SG + 2 SF). In return, this way to consider player roles constitutes a simplification of player characteristics, as it is not accounting for the ability for some players to play in several tactical positions. This also introduces potential effects of role distribution within teams: in the present dataset, there are 16 PG, 21 SG, 28 SF, 8 PF, and 23 C.

The replication of the method conducted in Korte et al. (2019) to basketball data reveals greater differences of FC (*i.e.*, moderate effect compared to small effect) and comparable differences of FB (*i.e.*, small effects) between positions. This indicates that the metrics they proposed for football are also suitable for basketball, and that basketball roles are more distinct than football roles in their involvement in the organizational processes of the attacking play. However, unlike the original study, present work shows no significant differences between the involvement of the different positions and the outcome of possession. This is possibly due to radical differences in the internal logic of the two sports, but also to the methods used to classify possessions as successful or unsuccessful: they used a proxy for goals scored (*i.e.*, entering the finish zone) to classify a possession as successful, whereas in present study a possession is considered to be positive when it ended with an actual positive result for the offensive team (*i.e.*, successful shot and/or a foul provoked). If other outcome classification could be used in basketball – for example considering as successful a possession ending with an open shot (Lucey et al., 2014) – the present methodological choice is expected to be as close as possible of the actual team performance, probably making it more challenging to predict the outcome.

But interestingly, the same analyze conducted with the metrics adapted to graphlet-level reveals some differences. In particular, having PGs present on fewer graphlets and being more intermediate would benefit in the case of possessions starting with an offside ball, suggesting that reducing an initiation phase characterized by a SG carrying the ball for several seconds would be a more effective way of organizing team behavior. Conversely, limiting the importance SG in possessions starting with an inside ball could also lead to more positive outcomes. There are also some effects for PF and SF, pointing to the benefits of involving them less in the organizational process. More broadly, these results indicate that traditional flow-based metrics might overlook the complexity of player roles in comparison to the flow-based metrics adapted to graphlet-level. This difference between the two types of metrics is also highlighted by the nonlinear and non-proportional relationships between them, further emphasizing that they do not measure the exact same things.

*Player roles: the effect of time pressure*

Beyond the quantitative variations in player roles (*i.e.*, in terms of FC and FB), there are also qualitative changes that can be observed in the patterns of interaction within which the different tactical playing positions are involved according to the shot clock. The 3-phase dynamic observed at macro-level also apply at micro-level, for each tactical playing position: as with team behavior, player roles are more specialized when the shot clock value is high or low, and more flexible at intermediate values. The third hypothesis of this work can therefore be partially confirmed: a role specialization process is indeed observed as the time remaining to shot decreases. However, entropy levels also indicate that player roles are very specific at the beginning of a possession, meaning that teams also exhibit a more mechanistic team structuring – albeit in a different way. The initial phase is therefore characterized by the setting up of a flexible organization, out of an organization that was more specialized at ball recovery. This flexible organization, probably relevant to provide some shooting opportunities, changes back to a more specialized organization again when time pressure becomes critically high. This specialization process is consistent with the theoretical use of a more mechanistic team structuring to cope with time pressure, and it may also be understood as the expression of *temporal leadership* (*i.e.*, leader behaviors that aid in structuring, coordinating, and managing the pacing of task accomplishment within teams) as a strategy to cope with time pressure (Liu & Liu, 2018).

Although this is a shared dynamic, there are still some differences between positions: PF and C roles appear to be more specialized than PG, SG and SF during the initial phase, and their peak flexibility is reached for lower shot clock values. Similarly, the way the possession started has an impact on player roles: possessions starting with an offside ball produce a much more specialized role for PG in the first moments of a possession, and seem to induce all positions to maintain a high level of flexibility for longer. These findings emphasize that role specialization is not only a function of time pressure, but also results of an interaction between the role itself and the tactical situation.

*Implications, limitations and perspectives*

This study reveals some insights into the functioning and performance of basketball teams during set plays characterized by a limited time, at both a meso and micro levels of analysis. In particular, this work revealed that after an initial phase during which teams move from a mechanistic structuring (*i.e.*, with specialized roles) to set up a more organic one (*i.e.*, with flexible roles), basketball teams maintain and try to exploit this organization until team member roles get more specific again through the effect of time pressure. If the findings are rooted in the context of basketball teams, some of them might be extended to other team settings where time pressure is a determinant factor (*e.g.*, medical teams during emergencies, work groups with deadlines), requiring teams to dynamically (re)organize their structure in order to adapt to this specific constraint.

Although some of the results specific to basketball may appear rather obvious (*e.g.*, PG has a very stereotyped role consisting of carrying the ball into opponent's half on possessions starting with an offside ball), finding them still underlines the relevance of the model defined for studying basketball teams. This is further supported by the fact that the model was also sensitive to initial conditions (*i.e.*, the way possessions started), which in turn suggests that other constraints could be integrated to study their effects (*e.g.*, defensive pressure, current score, the players on the court). In fact, beyond the results themselves, this work constitutes a new methodological framework that can be useful in various contexts where an adaptive complex system perspective can be applied. By providing a way to capture the dynamics of interactions within network analysis, the method used offers a more accurate representation of system's (re)organization processes and its temporal adaptations. Just as the findings demonstrated that the metrics used in Korte et al. (2019) – initially designed to evaluate player roles in football teams – are also useful in the case of basketball teams, the proposed adapted metrics are expected to be useful to analyze individual roles in other social networks. Similarly, the entropy-based metric – used to reflect the diversity of the temporal patterns of interaction among individuals within a team or the diversity of the temporal patterns in which a given player is involved – is also suitable for the analysis of other team settings. This combined approach integrating analyses at micro and meso levels aims more broadly to contribute to the SNA toolbox – particularly in contexts in which time is an important factor shaping the interactions.

However, some limitations should be considered. First of all, the dataset remains limited, meaning that the results should be interpreted with caution when considering their generalization to basketball team behavior in general – and even more in the case of other team settings. In particular, such a study would need to be conducted on a larger dataset to include other tactical situations (*e.g.*, when the ball is recovered in the attacking zone) in order to further investigate the interaction effect of time pressure and of tactical situations. Otherwise, for a more in-depth analysis of time pressure, it would be appropriate to conduct an experiment in which the tactical situations and the time remaining on the shot clock are controlled. Also, other methodological choices could be made regarding roles definition and possession outcomes classification. Indeed, this work was supported by the traditional definitions of playing positions in basketball (*i.e.*, PG, SG, SF, PF, C), which might have lost some of their relevance in modern basketball. Moreover, the outcome was classified as either positive or negative, but future deepen research could also consider performance outcomes as a quantitative variable such as the points per possession, a qualitative variable such as getting an open shot, or as a more complex model such as expected points (like expected goals in football). Finally, a team-by-team analysis would be beneficial for a deeper understanding of team (re)organization processes and role repartitions, allowing for the consideration of team-specific and/or individual-specific adaptation strategies.

**Conclusion**

In this work, basketball offensive teams were conceptualized as complex adaptive systems to analyze their (re)organizational processes according to the time remaining to shoot in different tactical situations. To do so, teams were modeled using temporal passing networks, and several metrics grounded in social network analysis were calculated at the meso and micro levels. Thanks to a finer integration of the temporal dimension, it was possible to focus on the different temporal patterns of interaction used by teams and on the role of the players within those patterns. As a result, a 3-phase dynamic has been identified at team level, and although it applies to all playing positions, the study of player roles has revealed quantitative and qualitative changes – some of which linked to team performance.

While the findings of this work are grounded in the context of basketball, their relevance should be extensible to other team sports and, beyond sports, to other team contexts in which time matters in the understanding of interactions between members.

This work can provide a useful tool for studying a wide range of social networks, by offering a multi-level solution particularly suitable for studying the dynamical (re)organizational processes of a system. Therefore, this work is expected to contribute to the existing scientific literature on network analysis applied to real data, with potential applications in both sport sciences and organizational studies.

Disclosure of interest
The authors report there are no competing interests to declare.

Data availability
The dataset used in this study is available in an OSF repository at https://osf.io/xv73a/. Data at different stages of processing are available from the corresponding author, on reasonable request.

Fundings
The work of Quentin Bourgeais and Ludovic Seifert was partially supported by the French ANR, project ANR-19-STHP-0006 (TEAM SPORTS). The work of Eric Sanlaville was partially supported by the French ANR, project ANR-22-CE48-0001 (TEMPOGRAL). The work of all authors was also partially supported by the French ANR, project ANR-23-CE38-0008 (DYNATEAM).

Authors contribution
Conceptualization: QB, RC, ES, LS; Methodology: QB, RC, ES; Data curation and formal analysis: QB; Writing - original draft: QB; Writing - review & editing: QB, RC, ES, LS; Supervision: RC, ES, LS; Funding acquisition: LS.